\documentclass[a4paper]{article}

\usepackage{itgspeech2023}    
\usepackage{times} 
\usepackage[english]{babel}   
\usepackage[ansinew]{inputenc}
\usepackage[T1]{fontenc}      
\usepackage[sort&compress,numbers]{natbib}	
\usepackage{amsmath,amssymb}
\usepackage{graphicx}
\usepackage[colorlinks=false,pdfborder={0 0 0}]{hyperref}
\usepackage{units}

\usepackage{color, colortbl}
\usepackage{array,multirow,graphicx}
\usepackage{tikz}
\usetikzlibrary{decorations.pathreplacing}

\usepackage{booktabs}
\usepackage{float}
\usepackage{microtype}

\definecolor{Gray}{gray}{0.9}
\definecolor{Gray_2}{gray}{0.75}
\definecolor{Gray_3}{gray}{0.98}

\title{Long-term Conversation Analysis: Exploring Utility and Privacy}

\author{Francesco Nespoli$^{1,2,*}$, Jule Pohlhausen$^{3,*}$, Patrick A. Naylor$^{2}$, Joerg Bitzer$^{3,4}$}

\address{$^1$Microsoft, London, UK\\
$^2$Dept. Electrical and Electronic Engineering, Imperial College London\\
$^3$Institute of Hearing Technology and Audiology, Jade University of Applied Science, Oldenburg, Germany\\
$^4$Fraunhofer IDMT Dept. HSA, Oldenburg, Germany\\
$^*$Equal contribution\\
Email: \footnotesize\texttt{\{jule.pohlhausen,joerg.bitzer\}@jade-hs.de; fnespoli@microsoft.com; p.naylor@imperial.ac.uk}
}

\begin{document}

\maketitle

\begin{abstract} 
The analysis of conversations recorded in everyday life requires privacy protection. In this contribution, we explore a privacy-preserving feature extraction method based on input feature dimension reduction, spectral smoothing and the low-cost speaker anonymization technique based on McAdams coefficient. We assess the utility of the feature extraction methods with a voice activity detection and a speaker diarization system, while privacy protection is determined with a speech recognition and a speaker verification model. We show that the combination of Mc\-Adams coefficient and spectral smoothing maintains the utility while improving privacy.

\end{abstract}

\section{Introduction}


Analysing conversations in everyday life situations is of great interest for many diverse research fields. Examples range from measurements of children's language-learning environments \cite{vandam_quantity_2012} to social interaction analysis of people \cite{heritage2008conversation}, e.g., in the case of patients affected by a mental disorder or hearing impairments \cite{smeds_estimation_2015}. In this context, we aim to analyse long-term conversations from participants with dementia, recorded with portable devices over multiple days in their everyday life. Besides measuring how many conversations individuals do have in a day, it is important to explore other characteristic aspects such as duration of each encounter, the number and variety of communication partners and speaker turns \cite{segerdahl1998CA}. However, due to the fact that speech includes important personal identifiable information (PII) both in the semantic and acoustic domain and because everyday-life includes unconstrained and unpredict\-able situations both in private and public, audio recordings in these scenarios raise privacy concerns. Examples of semantic PII include full names, security numbers and geographical position. Moreover, acoustic features extracted from the voice such as prosody, speaking rate, accent and intonation enclose a variety of PII such as personality, physical characteristics, emotional state, age and gender that can be identified \cite{Krger2020} and therefore employed for privacy attacks. Because of these reasons, data protection regulation such as the European Union General Data Protection Regulation (EU GDPR) \cite{EUGDPR} enforces privacy preservation solutions for speech data. 

Pioneering acoustic privacy protection approaches explored several research directions such as computing priva\-cy-preserving features \cite{six}, working with encrypted speech signals \cite{seven}, learning adversarial features \cite{nine}, or performing federated learning \cite{federated}. However, the aforementioned feature or model-level privacy protection techniques can not be applied in our scenario mainly due to the low computational power of portable recording devices and their power consumption constraints. In \cite{wyatt2007conversation}, the authors proposed a conversation detection and speaker diarization system using low-cost privacy-preserving features with no possibility of linguistic content reconstruction. However, the system needs one audio stream per-speaker and accesses a central node combining the information from all the streams, which is not feasible in our application. Another low-cost solution to these problems is to limit the recordings to privacy-preserving acoustic features and conduct offline analyses \cite{bitzer2016privacy}. 

Following this idea, in this contribution we investigate the possibility of combining the smoothed and subsampled power spectral densities (PSD) \cite{bitzer2016privacy} to protect the linguistic PII and a lightweight anonymization technique, based on the McAdams coefficient \cite{patino2021mcadams}, to protect the acoustic PII. Moreover, we consider attack and trust models that rely on the same feature extraction process. Therefore, we can decrease the feature resolution in the time and/or frequency domain, and directly observe the impact on the utility as well as on the privacy. The utility assessment considers the Matthews correlation coefficient (MCC) of a Voice Activity detection (VAD) model and the Diarization Error Rate (DER) extracted from a Speaker Diarization (SD) system. This choice was determined by the fact that conversation analysis requires to first detect voiced time segments and then attribute to each of the segments a unique speaker label, which is not necessarily linked to the speaker's identity. Furthermore, the privacy assessment consists of two parts: the speech content part
carried out with an Automatic Speech Recognition (ASR) model and the speaker identity part
performed with an Automatic Speaker Verification (ASV) system. The fundamental idea behind privacy preservation is to preserve utility while enforcing privacy, therefore minimizing DER while maximizing MCC, Word
Error Rate (WER) and Equal Error Rate (EER).
\begin{figure}[h]
	\centerline{\includegraphics[width=\columnwidth]{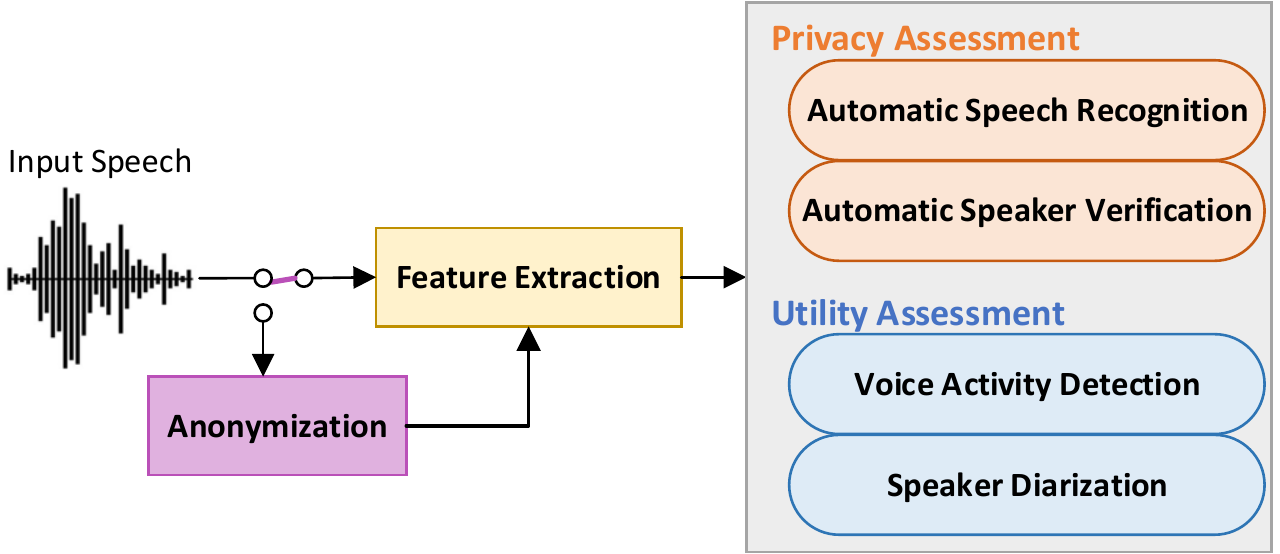}}
	\caption{Privacy and utility assessment. \textit{Anonymization} is an optional preprocessing stage.}
	\label{fig:pipeline}%
\end{figure} 
Figure~\ref{fig:pipeline} summarizes the privacy and utility assessment we employed in our experiments. 
To foster replicability, we made the code available online\footnote{
Soon available at \scriptsize{\url{https://github.com/ol-MEGA/ppca.git}}}.

\section{Framework description} 
\label{sec:framework}%

We conducted our experiments with the open-source speech-processing toolkit SpeechBrain \cite{speechbrain}. All models rely on the same feature extraction process, namely standard log Mel filterbank energies computed on time segments with a window size of \unit[25]{ms} and a hop size of \unit[10]{ms}. In the case of smoothed and subsampled PSD, the window size is \unit[25]{ms} with a hop size of \unit[$L = 12.5$]{ms}. Moreover, segments are smoothed with a first-order recursive filter with a smoothing time constant \unit[$\tau = 125$]{ms} and subsampled by a factor of $\tau/L = 10$. We use these smoothed and subsampled PSD in combination with the standard Mel filterbank to compute log Mel filterbank energies. To restore the original time resolution of the segments, we repeated the smoothed and subsampled PSD by the subsampling factor before applying the Mel filterbank and we refer to this feature as \textit{olMEGA}, since the features were developed for an open-source Ecological Momentary Assessment system \cite{Kowalk2020olMEGA} with the same name.

\subsection{Privacy Assessment}
In this contribution, we consider the two main privacy aspects for speech recordings \cite{nelus2016towards}: the linguistic speech content and the speaker identity. The following sections describe our ASR and ASV models.

\subsubsection{Automatic Speech Recognition}\label{sec:ASR}
The ASR system is based on a transformer acoustic model encoder and a joint transformer decoder with connectionist temporal classification (CTC) \cite{CTC}, with the decoding stage integrating also CTC probabilities. The privacy evaluation of the linguistic content relies on the WER metric which is calculated from the ASR output transcription as \begin{equation} \mathrm{WER} = \frac{N_{\mathrm{sub}}+N_{\mathrm{ins}}+N_{\mathrm{del}}}{N_{\mathrm{tok}}} \end{equation} where $N_{\mathrm{sub}},N_{\mathrm{ins}},N_{\mathrm{del}}$ are the number of substitutions, insertions and deletions in the ASR output and $N_{\mathrm{tok}}$ is the number of tokens in the ground-truth transcript.

\subsubsection{Automatic Speaker Verification}\label{sec:ASV}
In our experiments, the privacy attacker is \textit{semi-informed}, i.e., the strongest attack paradigm considered in \cite{voicepriv2020}. Semi-informed attackers have complete knowledge about the anony\-mization system. However, they do not have access to the specific parameters mapping original and anonymized speech. In this scenario, the attacker can anonymize a publicly available speech corpus and use it to fine-tune the ASV model therefore improving the recognition performance of the system on anonymized speech. The speaker verification model is based on the ECAPA-TDNN speaker encoder \cite{ECAPA} coupled with a simple cosine similarity scoring for the verification task. The privacy evaluation relies on the EER: given a generic biometric authentication system $\mathrm{G}$, $R_\mathrm{fa}^\mathrm{G}(\theta)$ and $R_\mathrm{fr}^\mathrm{G}(\theta)$ are the false acceptance and false rejection rates at a given decision threshold $\theta$. The EER corresponds to the rate at which $R_\mathrm{fa}^\mathrm{G}(\theta)=R_\mathrm{fr}^\mathrm{G}(\theta)$. 

\subsection{Utility Assessment}
The objective analysis of conversations requires a precise and robust detection of time segments containing speech. Furthermore, the speech segments need to be attributed to the corresponding speakers. The following sections describe the VAD and SD models.

\subsubsection{Voice Activity Detection}\label{sec:VAD}
The VAD \cite{speechbrain}, is based on a Convolutional Recurrent Deep Neural Network (CRDNN) architecture. The model computes the frame-level posterior probabilities which are then processed with a sigmoid function to perform binary classification (speech vs. non-speech). Utility evaluation relies on the \small \begin{equation} \mathrm{MCC} = \frac{\mathrm{TP}\times \mathrm{TN} - \mathrm{FP}\times \mathrm{FN}}{\sqrt{(\mathrm{TP+FP})\cdot (\mathrm{TP+FN})\cdot (\mathrm{TN+FP})\cdot (\mathrm{TN+FN})}} \end{equation}  \normalsize where TP, TN, FP and FN are the number of true positives, true negatives, false positives and false negatives, respectively.

\subsubsection{Speaker Diarization}\label{sec:SD}
The SD model is based on ECAPA-TDNN embeddings, the same employed by the ASV model, and performs Spectral Clustering (SC) to assign a relative speaker label to each time segment \cite{dawalatabad_ecapa-tdnn_2021}. The model relies on the oracle VAD information. We evaluate the SD utility with the \begin{equation} \mathrm{DER} = \frac{\mathrm{FA+MISS+ERROR}}{\mathrm{TOTAL}} \end{equation} where TOTAL is the total reference speaker time, FA is the total speaker time not attributed to the reference speaker, MISS is the total reference speaker time not attributed to the speaker and ERROR is the total reference speaker time attributed to the wrong speaker \cite{fiscus2006}. During DER evaluation a forgiveness collar of \unit[250]{ms} is used and the speaker overlap regions are included. This \textit{fair} setup \cite{Landini2022} takes into account that no realistic human annotation can achieve frame-level precision.

\section{Proposed Method} 
\label{sec:methods}%

Human experiments with mosaic speech - analogous to pixelated images - show a decreased speech recognition due to a degraded resolution in the time and/or frequency domain \cite{nakajima2018mosaic_speech}. Analogously, the performance of ASR models on mosaicized speech shows similar trends \cite{Adolfi2023}. 
Hence, our first experiment towards privacy preservation was to decrease the size of the Mel filterbank while covering the same frequency range. Secondly, we applied the olMEGA smoothing and subsampling \cite{bitzer2016privacy} with repetition before applying the Mel filterbank to further improve linguistic PII obfuscation. In addition, to further enforce privacy protection in the acoustic domanin, we applied a low-cost signal processing technique for anonymization. The McAdams coefficient \cite{patino2021mcadams} shifts the pole positions derived from the linear predictive coding (LPC) analysis of speech signals. However, it has been shown that anonymization using the McAdams coeffient in the condition of a semi-informed attacker (described in Section~2.1.2), leads to only marginally deteriorated WER \cite{voice2022, patino2021mcadams}. While utterance-wise McAdams anonymization leads to strong privacy preservation in terms of EER \cite{patino2021mcadams}, speaker-wise McAdams anonymization only modestly deteriorates EER \cite{voice2022}. Therefore, to further investigate McAdams anonymization capabilities, we combined McAdams coefficient with the olMEGA smoothing and subsampling \cite{bitzer2016privacy} in yet another condition explained in Section \ref{sec:Data}.


\begin{table*}[ht!]
    \centering
    

    \definecolor{Gray}{gray}{0.9}
    \definecolor{Gray_2}{gray}{0.75}
    \begin{tabular}{ccccccc|ccc|ccc}
    \toprule
    \textbf{Set}&\textbf{Gender}&\textbf{Weight}&&&&\multicolumn{6}{c}{\textbf{EER[\%]}} \\
    \midrule
    &&&\multicolumn{1}{c}{5}&\multicolumn{1}{c}{10}&\multicolumn{1}{c}{20}&\multicolumn{1}{c}{40}&\multicolumn{3}{c}{80}&\multicolumn{3}{c}{olMEGA} \tikz [remember picture] \node (rightmark) {}; \\
    \midrule
    &&&O&O&O&O&O&AN&FT&O&AN&FT \\
    \rowcolor{Gray}
    \rowcolor{Gray}
    Libri-dev    & female  & 0.25 & 7.53 & 3.97 & 2.98 & 3.12 & 2.70 & 9.81 & 13.58 & 5.97 & 14.21 & 15.19 \\
    \rowcolor{Gray}
                 & male    & 0.25 & 2.64 & 0.78 & 0.61 & 0.31 & 0.34 & 6.83 & 4.53 & 1.55 & 8.10 & 7.62 \\
    \rowcolor{Gray}
    VCTK-dev     & female  & 0.20 & 5.00 & 1.91 & 0.62 & 0.17 & 0.44 & 8.25 & 2.87 & 1.8 & 9.04 & 5.73 \\
    \rowcolor{Gray}
    
    diff.        & male    & 0.20 & 4.26 & 0.99 & 0.30 & 0.10 & 0.45 & 5.95 & 3.33 & 1.29 & 8.14 & 5.46 \\
    \rowcolor{Gray}
    VCTK-dev     & female  & 0.05 & 2.62 & 1.75 & 0.59 & 0.87 & 0.51 & 6.90 & 3.22 & 0.94 & 8.15 & 6.4 \\
    \rowcolor{Gray}
        comm.    & male    & 0.05 & 2.85 & 1.08 & 0.02 & 0.01 & 0.02 & 2.56 & 3.96 & 0.57 & 4.56 & 5.95 \\
    
    \midrule
    $Avg^W dev$ & &               & 4.67 & 1.88 & 1.1 & 0.96 & 0.97 & 7.47 & 6.12 & 2.57 & 9.65 & 8.56 \\
    \midrule
    
    \rowcolor{Gray_2}
    Libri-test   & female  & 0.25 & 4.57 & 1.32 & 0.54 & 0.69 & 0.37 & 8.76 & 9.35 &  2.19 & 14.57 & 10.95 \\
    \rowcolor{Gray_2}
                 & male    & 0.25 & 3.79 & 1.77 & 0.72 & 0.45 & 0.89 & 6.64 & 3.10 & 1.55 & 8.02 & 6.9 \\
    \rowcolor{Gray_2}
    VCTK-test    & female  & 0.20 & 9.98 & 3.35 & 1.49 & 0.98 & 0.93 & 9.41 & 4.68 & 2.74 & 11.53 & 10.09 \\
    \rowcolor{Gray_2}
       diff.     & male    & 0.20 & 4.72 & 1.43 & 0.63 & 0.36 & 0.41 & 11.20 & 3.26 & 1.43 & 8.67 & 4.93 \\
    \rowcolor{Gray_2}
    VCTK-test    & female  & 0.05 & 3.47 & 1.42 & 0.56 & 0.58 & 0.82 & 8.72 & 2.25 & 2.02 & 12.43 & 6.07 \\
    \rowcolor{Gray_2}
       comm.     & male    & 0.05 & 1.95 & 0.28 & 0.01 & 0.02 & 0.26 & 4.81 & 2.83 & 0.06 & 4.46 & 3.73 \\
    
    \midrule
    $Avg^W test$ & &  &              5.30 & 2.14 & 0.77  & 0.58  & 0.64  & 8.65 & 4.95 & 1.87 & 10.53 & 7.96  \\
    \bottomrule
    \end{tabular}
    \caption{ASV performance on the Voice Privacy Challenge \cite{voicepriv2020} dataset for different Mel feature dimensions, olMEGA and anonymized speech. O: original data. AN: anonymized data and model trained on unprocessed speech. FT: anonymized data and model fine-tuned on anonymized speech.}
    \label{tab:ASV}
    \end{table*}
    
\section{Experiments and Results}
\label{sec:experiments}%

This Section describes the datasets used in this contribution, the privacy assessment results from ASR and ASV, and the utility assessment results from VAD and SD.

\subsection{Data}\label{sec:Data}
In all our experiments we used publicly available datasets sampled at 16~kHz. The ASR models were trained on the full 960 hours of LibriSpeech \cite{librispeech} and tested on the test-clean and test-others subsets. For speaker verification, all models were trained on VoxCeleb~2 \cite{VoxCeleb2}. In the case of ASR and ASV models fine-tuned an anonymized data, we used an anonymized version of LibriSpeech-360 obtained by randomly sampling a McAdams coefficient from $(0.5, 0.9)$ for each utterance. Finally, the ASV was evaluated on the same splits as in \cite{voicepriv2020} in both of the cases of clean and anonymized speech. In this case, the anonymized splits were obtained by applying a random McAdams coefficient to each simulated meeting based on the Voice Privacy 2020 Challenge (VPC) \cite{voicepriv2020} dataset described below. The VAD models were trained and tested on the train and test split of LibriParty \cite{speechbrain}, respectively. In the case of anonymized speech, the aforementioned splits were anony\-mized following the same strategy applied to LibriSpeech-360. Further, SD performance was evaluated on the Augmented Multi-party Interaction (AMI) Meeting Corpus \cite{carletta_unleashing_2007} with the standard \textit{Full-corpus-ASR} partition using the \textit{HeadsetMix} recording streams. The oracle VAD information was extracted from the manual ground truth annotations.

\subsubsection{Simulations}\label{sec:sims}

In order to show ASV and SD results on the same data, we simulated conversations based on the VPC \cite{voicepriv2020} test and evaluation sets. Using the Multi-purpose Multi-Speaker Mixture Signal Generator (MMS-MSG) \cite{cord_mms-msg_2022} we generated meeting-like speech with 3-4 speakers per meeting and no overlapping speech. First, the silences at the beginning and end of each utterance were removed. To produce a fine-grained VAD, we used the VAD described in Section~\ref{sec:VAD} and applied an energy-based threshold on the detected speech segments to further improve the resolution of on- and off-sets. Second, the speakers were randomly assigned to a meeting. We employed the activity-based speaker turn sampling proposed by \cite{cord_mms-msg_2022} and randomly sample the utterances per-speaker until exhaustion. We utilized the complete VPC test and evaluation data without repetitions. We refer to this data as VPC simulated meetings.

\subsection{Automatic Speech Recognition}\label{sec:resultsASR}
Table~\ref{tab:ASR} reports the WER of the models trained for 90 epochs with a batch size of 256 for each input feature. Anonymized speech was decoded with the model trained on clean data and with a second model, fine-tuned for 30 epochs on the anonymized LibriSpeech-360 described in Section~\ref{sec:Data}. 

\begin{table}[H]
\centering


\definecolor{Gray}{gray}{0.9}
\definecolor{Gray_2}{gray}{0.75}

\begin{tabular}{ccccc}
\toprule
\textbf{Set}&\multicolumn{4}{c}{\textbf{WER[\%]}} \tikz [remember picture] \node (rightmark) {}; \\
\midrule
&20&40&80&olMEGA\\ 
\midrule
\rowcolor{Gray}
\rowcolor{Gray}
test-clean    & 3.56  & 2.36 & 2.34 & 5.97   \\
\rowcolor{Gray}
test-other    & 8.97 & 5.46 & 5.55 & 16.22  \\

\midrule

\rowcolor{Gray_2}
test-clean-anon     & 18.42  & 9.10 & 11.40 & 93.90  \\
\quad + FT & 5.70    & 3.03 & 2.96 & 11.66  \\
\rowcolor{Gray_2}
test-other-anon     & 52.12  & 30.02 & 35.91 & 102.72  \\
\quad + FT & 17.31    & 8.48 & 8.04 & 32.95  \\

\bottomrule
\end{tabular}
\caption{ASR decoding on test subsets of \cite{librispeech}
for different Mel input feature dimensions, olMEGA and anonymized speech with models trained on clean (dark gray rows) and fine-tuned (FT) on anonymized (white rows) data.}    
\label{tab:ASR}
\end{table}
\vspace*{-10pt}

\begin{table*}[ht!]
    \centering
    \begin{tabular}{ccccc|ccc|ccc}
    \toprule
    \textbf{Set} & \multicolumn{10}{c}{\textbf{DER[\%]}} \tikz [remember picture] \node (rightmark) {}; \\
    \midrule
    & 5 & \multicolumn{1}{c}{10} & \multicolumn{1}{c}{20} & \multicolumn{1}{c}{40} & \multicolumn{3}{c}{80} & \multicolumn{3}{c}{olMEGA} \tikz [remember picture] \node (rightmark) {}; \\
    \midrule
    &O&O&O&O&O&AN&FT&O&AN&FT \\
    \rowcolor{Gray_3}
    AMI-dev    & 18.70 & 17.15   & 16.45 & 15.92 & 16.22 & 28.42 & 18.34 & 15.71 & 33.35 & 25.31 \\
    \rowcolor{Gray_3}
    AMI-eval   & 18.38 &  16.98  & 16.71 & 16.59 & 16.32 & 29.13 & 19.91 & 17.12 & 30.55 & 24.86 \\
    \midrule
    \rowcolor{Gray}
    Libri-dev  & 6.89  & 5.25    &  7.79 & 10.55 &  7.54 &  5.80 &  9.69 &  5.16 &  6.70 & 5.91 \\
    \rowcolor{Gray}
    VCTK-dev   & 0.05  & $<0.01$ &  $<0.01$  & $<0.01$   & 0.86  &  1.17 &  0.08 &  0.03 &  5.05 & 0.71 \\
    \midrule
    \rowcolor{Gray_2}
    Libri-test & 5.26  & 6.46    &  8.40 &  5.09 &  6.30 &  8.64 &  16.1 &  3.23 &  9.31 & 9.81 \\
    \rowcolor{Gray_2}
    VCTK-test  & 0.12  & 0.01    &  1.44 &  1.39 &  0.63 & 11.87 &  0.67 &  0.33 & 15.14 & 1.23 \\
    \bottomrule
    \end{tabular}
\caption{DER on the AMI corpus development and evaluation sets and the VPC simulated meetings sets for different Mel input feature dimensions, olMEGA and anonymized speech. The number of speakers per meeting is estimated and a forgiveness collar of \unit[250]{ms} is applied. O: original data. AN: anonymized data and model trained on unprocessed speech. FT: anonymized data and model fine-tuned on anonymized speech.}
\label{tab:DER}
\end{table*}

\subsection{Automatic Speaker Verification}\label{sec:resultsASV}
Table~\ref{tab:ASV} shows the EER for the models trained on unprocessed and anonymized data. The models were trained for 30 epochs with a batch size of 512 on clean data. We used the anonymized LibriSpeech-360 described in Section~\ref{sec:Data} to fine-tune the models trained on VoxCeleb~2 on anonymized data for 30 more epochs, with the idea of minimizing the mismatch between training and testing sets.

\subsection{Voice Activity Detection}\label{sec:resultsVAD}
The VAD models were trained on unprocessed and anony\-mized LibriParty data for 100 epochs with a batch size of 2. For models tested and trained on unprocessed data, the MCC ranged between 0.70 and 0.86 with 5 and 80 Mels as input feature dimension, respectively. The model trained with preprocessing olMEGA and 80 Mels shows a MCC of 0.84 and 0.82, trained and tested on unprocessed and anonymized data, respectively.


\subsection{Speaker Diarization}\label{sec:resultsSD}
The SD model employs the same ECAPA-TDNN embedding as used for the ASV in Section~\ref{sec:resultsASV}. The pruning threshold for the affinity matrix during SC is determined on the AMI and Libri development set, respectively.
Table~\ref{tab:DER} summarizes the SD performance on the AMI corpus development and evaluation sets and the VPC simulated meetings. The number of speakers per meeting is estimated and limited to ten.

\section{Discussion}
The reduction of the input Mel filterbank leads to degraded results for the ASR model (Table~\ref{tab:ASR}) in accordance with \cite{Adolfi2023}. Moreover, McAdams anonymization further harms the \-WER (Table~\ref{tab:ASR}, dark gray rows), which can be partially recovered by fine-tuning the model on anonymized speech (Table~\ref{tab:ASR}, white rows). However, ASV performance is more robust even for low (10 Mels, Table~\ref{tab:ASV}) dimensional inputs. Only for the 5 Mel condition we observe a consistent drop in performance. Similarly, the VAD performance is marginally affected by the decreased size of the Mel filterbank and only the 5 Mels condition leads to a degraded MCC. This result is consistent with state-of-the-art (SOTA) VAD algorithms relying on low-scaled energy information \cite{marzinzik2002vad}. However, the VAD is robust against the McAdams anony\-mization, especially if trained on anonymized data. Furthermore, the SD performance is robust for models trained with olMEGA (Table~\ref{tab:DER}: 80 Mels and olMEGA, O columns). Contrarily to the EER, the DER remains nearly constant with decreased Mel filterbank resolution. The DER on the VPC simulated meetings shows a higher variation compared to the AMI corpus. Morevover, the SD performance on the VPC simulated meetings with oracle speaker information is nearly perfect (results not shown). One possible explanation for these results could be related to the clean recording conditions of the original VPC data. Furthermore, the differences between the Libri and VCTK subsets can be explained by the overestimated number of speakers for the Libri subsets. 

Overall, the feature extraction with olMEGA leads to privacy improvements both in the acoustic and in the semantic domain. This is particularly evident when olMEGA is applied in combination with McAdams coefficient anony\-mization (Table~\ref{tab:ASV}: $Avg^W dev$ , $Avg^W test$. Table~\ref{tab:ASR}: test-clean-anon+FT, test-other-anon+FT). Furthermore, enhanc\-ed privacy protection comes with a gain (Table~\ref{tab:DER}: Libri-dev and Libri-test, olMEGA FT column) or only small decrements (Table~\ref{tab:DER}: VCTK-dev and VCTK-test, olMEGA FT column) in DER on the VPC simulated dataset when compared with standard 80 mel input features (Table~\ref{tab:ASR}, 80 Mels, FT columns). However, this conclusion doesn't directly extend to the AMI dataset and further investigations are needed before drawing any conclusive result.

\section{Conclusions}
This contribution presents an analysis of the trade-off between privacy and utility of different signal processing based feature extraction methods in combination with the McAdams speaker anonymization technique. We showed that reducing the input feature dimension in combination with spectral smoothing and the McAdams anonymization technique leads to improved privacy preservation in comparison to unprocessed signals while retaining high utility on SD and VAD on the VPC simulated meetings. This result however doesn't seem to generalize to the AMI dataset. Further research directions will include a more sophisticated speaker-aware diarization system in combination with a SOTA speaker anonymization technique. Moreover, besides the standard Mel filterbank energies, we will investigate new input feature combinations that jointly optimize utility and privacy.



\section{Acknowledgements}

This work was supported by the European Union Horizon 2020 program under the Marie Sklodowska-Curie grant No 95636 and by the Graduation program of Jade University of Applied Sciences (Jade2Pro 2.0).

\pagebreak	
\clearpage
\small
\bibliographystyle{ieeetr}
\bibliography{sapstrings,literature}

\end{document}